\documentclass[12pt]{article}
\usepackage{graphicx}
\begin{document}
\centerline{\bf Majority-vote on undirected Barab\'asi-Albert networks}

\bigskip
\centerline{F.W.S. Lima}
 
\bigskip
\noindent
Departamento de F\'{\i}sica, 
Universidade Federal do Piau\'{\i}, 57072-970 Teresina - PI, Brazil

\medskip
  e-mail: wel@ufpi.br
\bigskip
 
{\small Abstract: On Barab\'asi-Albert networks 
with $z$ neighbours selected by each added site, the Ising model was seen
to show a spontaneous magnetisation. This spontaneous magnetisation was
found below a critical temperature which increases logarithmically with
system size. On these networks the
majority-vote model with noise is now studied through Monte Carlo simulations.
However, in this model,  the order-disorder phase transition of the
order parameter is well defined in this system and this wasn't
found to increase logarithmically with
system size. We calculate the value of the critical 
noise parameter $q_{c}$ for several values of connectivity $z$ of the
undirected Barab\'asi-Albert network. The critical 
exponentes $\beta/\nu$, $\gamma/\nu$ and $1/\nu$ were calculated for
several values of $z$.}
 
 
\bigskip

 {\bf Introduction}
 
 It has been argued that nonequilibrium stochastic spin systems on 
 regular square lattice with up-down symmetry fall in the universality
 class of the equilibrium Ising model \cite{g}. This conjecture was
 found in several models that do not obey detailed balance \cite{C,J,M}.
 Campos $et$ $al$. \cite{campos} investigated the majority-vote model 
 on small-world network by rewiring the two-dimensional square lattice. These
 small-world networks, aside from presenting quenched disorder, also 
 possess long-range interactions. They found that the critical exponents 
 $\gamma/\nu$ and $ \beta/\nu$ are different from the Ising model and depend
 on the rewiring probability. However, it was not evident whether the
 exponent change was due to the disordered nature of the network or due to 
 the presence of long-range interactions. Lima $et$ $al$. \cite{lima0} 
 studied the majority-vote model on Voronoi-Delaunay random lattices 
 with periodic boundary conditions. These lattices posses natural quenched
 disorder in their conecctions. They showed that presence of quenched 
 connectivity disorder is enough to alter the exponents $\beta/\nu$
 and $\gamma/\nu$ the pure model and therefore that is a relevant term to
 such non-equilibrium phase-transition. 
 Sumour and Shabat \cite{sumour,sumourss} investigated Ising models on 
 directed Barab\'asi-Albert networks \cite{ba} with the usual Glauber
 dynamics.  No spontaneous magnetisation was 
 found, in contrast to the case of undirected  Barab\'asi-Albert networks
 \cite{alex,indekeu,bianconi} where a spontaneous magnetisation was
 found lower a critical temperature which increases logarithmically with
 system size. Lima and Stauffer \cite{lima} simulated
 directed square, cubic and hypercubic lattices in two to five dimensions
 with heat bath dynamics in order to separate the network effects  form
 the effects of directedness. They also compared different spin flip
 algorithms, including cluster flips \cite{wang}, for
 Ising-Barab\'asi-Albert networks. They found a freezing-in of the 
 magnetisation similar to  \cite{sumour,sumourss}, following an Arrhenius
 law at least in low dimensions. This lack of a spontaneous magnetisation
 (in the usual sense)
 is consistent with the fact
 that if on a directed lattice a spin $S_j$ influences spin $S_i$, then
 spin $S_i$ in turn does not influence $S_j$, 
 and there may be no well-defined total energy. Thus, they show that for
 the same  scale-free networks, different algorithms give different
 results.  More recently, Lima \cite{lima1} investigated the majority-vote model
 on directed 
 Barab\'asi-Albert network and calculated the
 $\beta/\nu$, $\gamma/\nu$, and $1/\nu$ exponents and  these was 
 different from the Ising model and depend on the values of
 connectivity $z$ of the 
 directed Barab\'asi-Albert network. Now, we calculate the same
 $\beta/\nu$, $\gamma/\nu$, and $1/\nu$ exponents for
 majority-vote model on {\it undirected} 
 Barab\'asi-Albert network and  these are 
 different from the Ising model and depend on the values of
 connectivity $z$ of the 
 directed Barab\'asi-Albert network.
 
\bigskip
 
\begin{figure}[hbt]
\begin{center}
\includegraphics[angle=-90,scale=0.46]{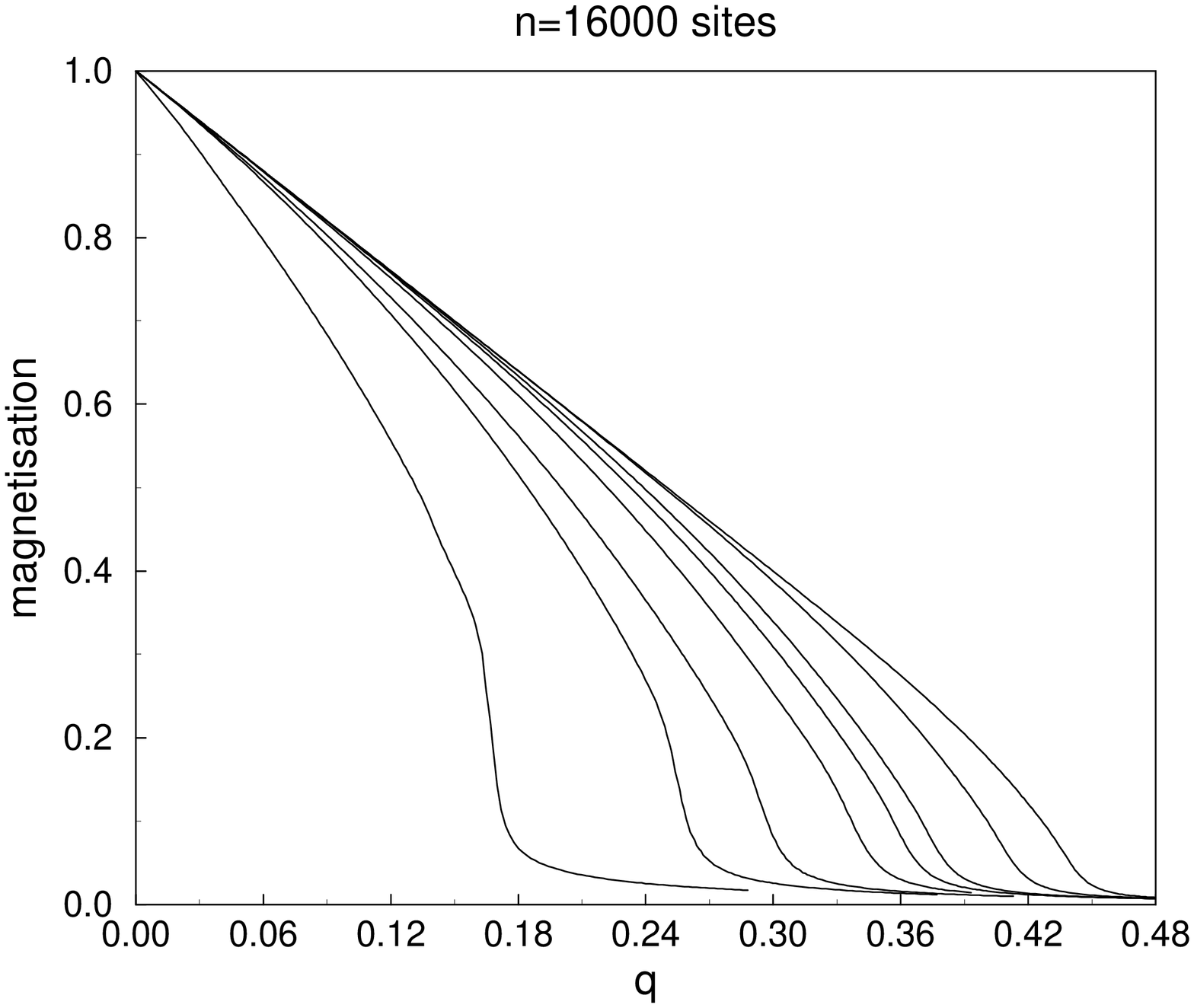}
\includegraphics[angle=-90,scale=0.46]{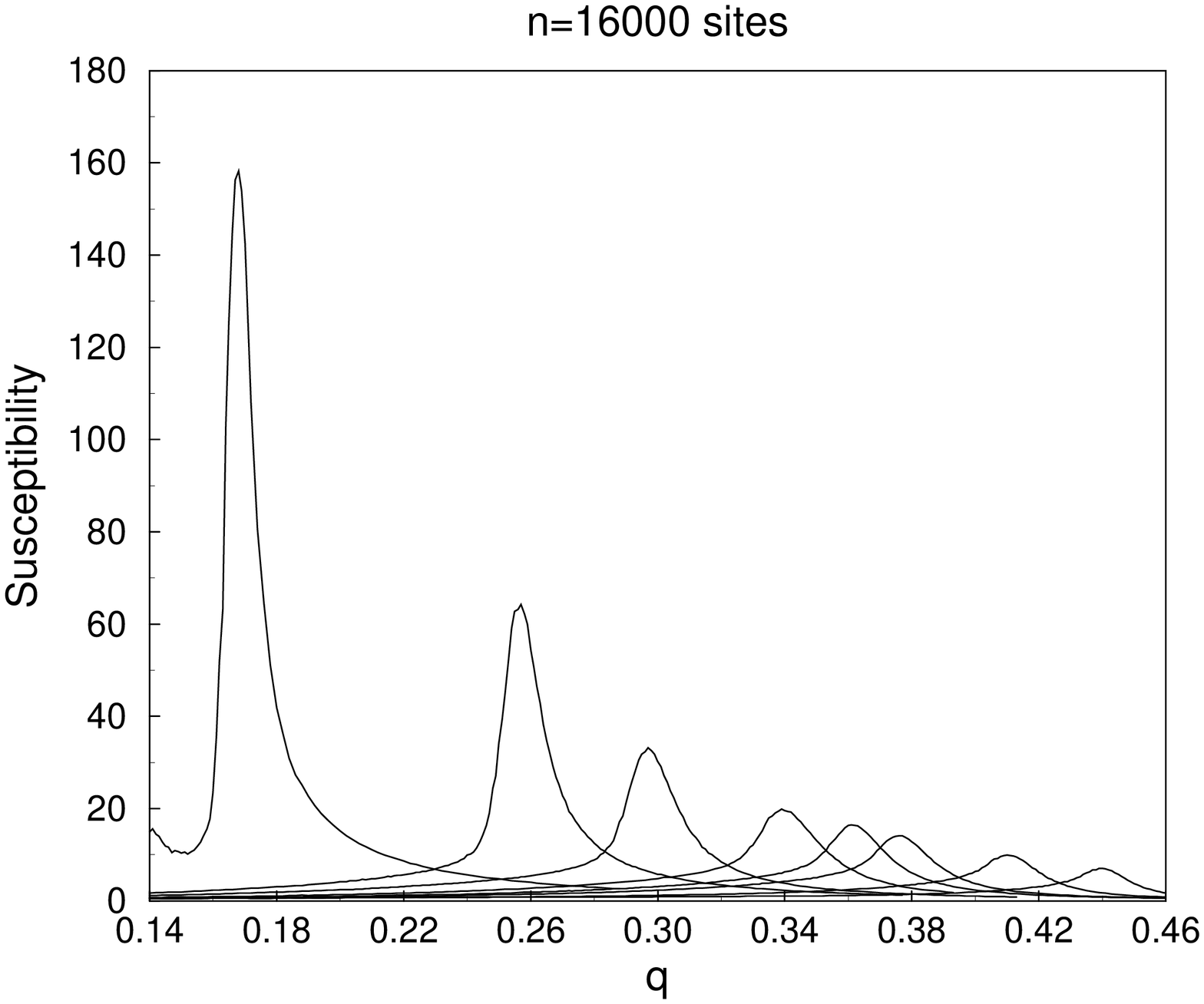}
\end{center}
\caption{
Magnetisation and susceptibility as a function of the noise parameter $q$, for
$N=16000$ sites. From left to right we have $z=2$, $3$, $4$, $6$, $8$, $10$, $20$,
and $50$ .}
\end{figure}
  
\bigskip
 
\begin{figure}[hbt]
\begin{center}
\includegraphics[angle=-90,scale=0.46]{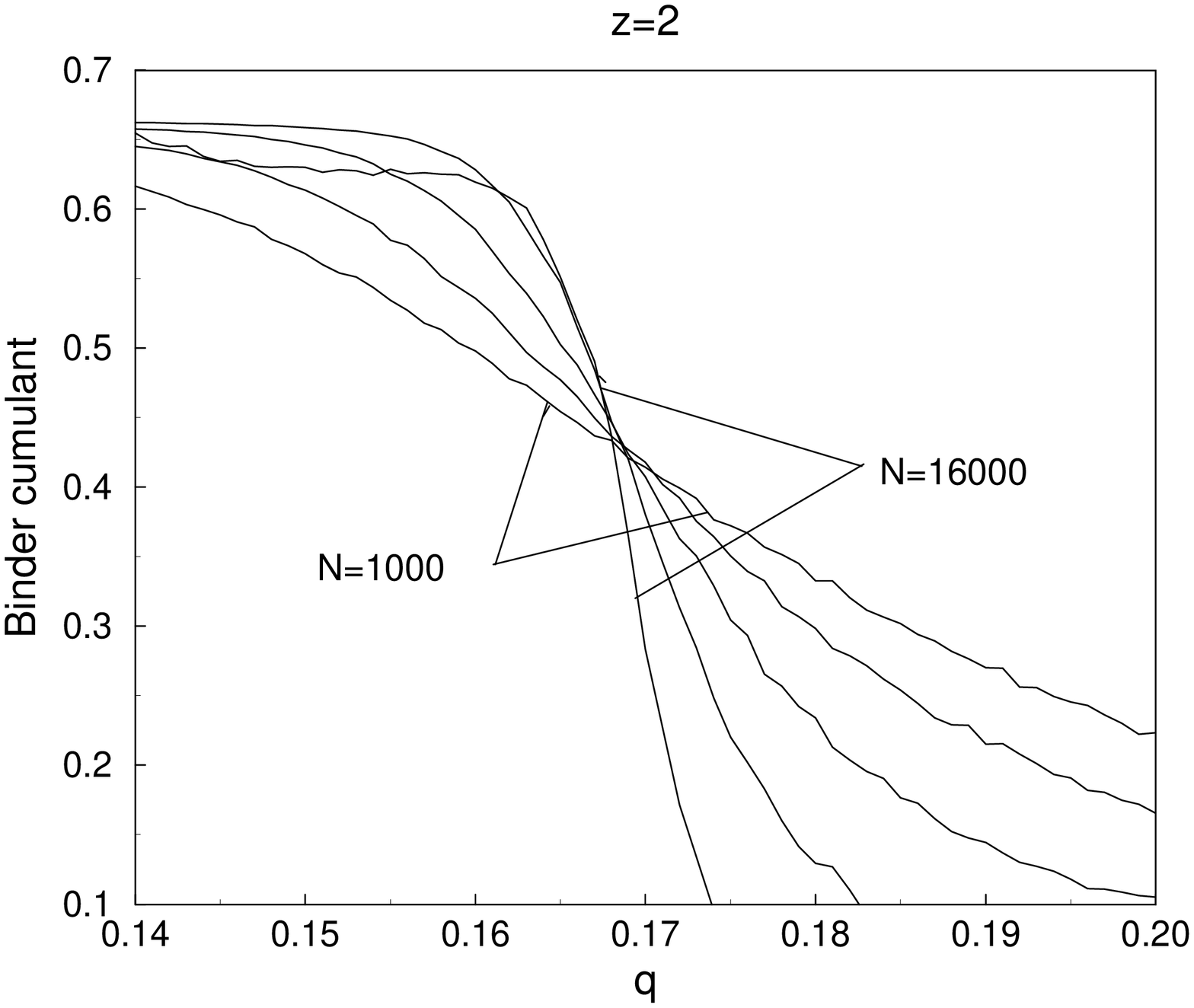}
\includegraphics[angle=-90,scale=0.46]{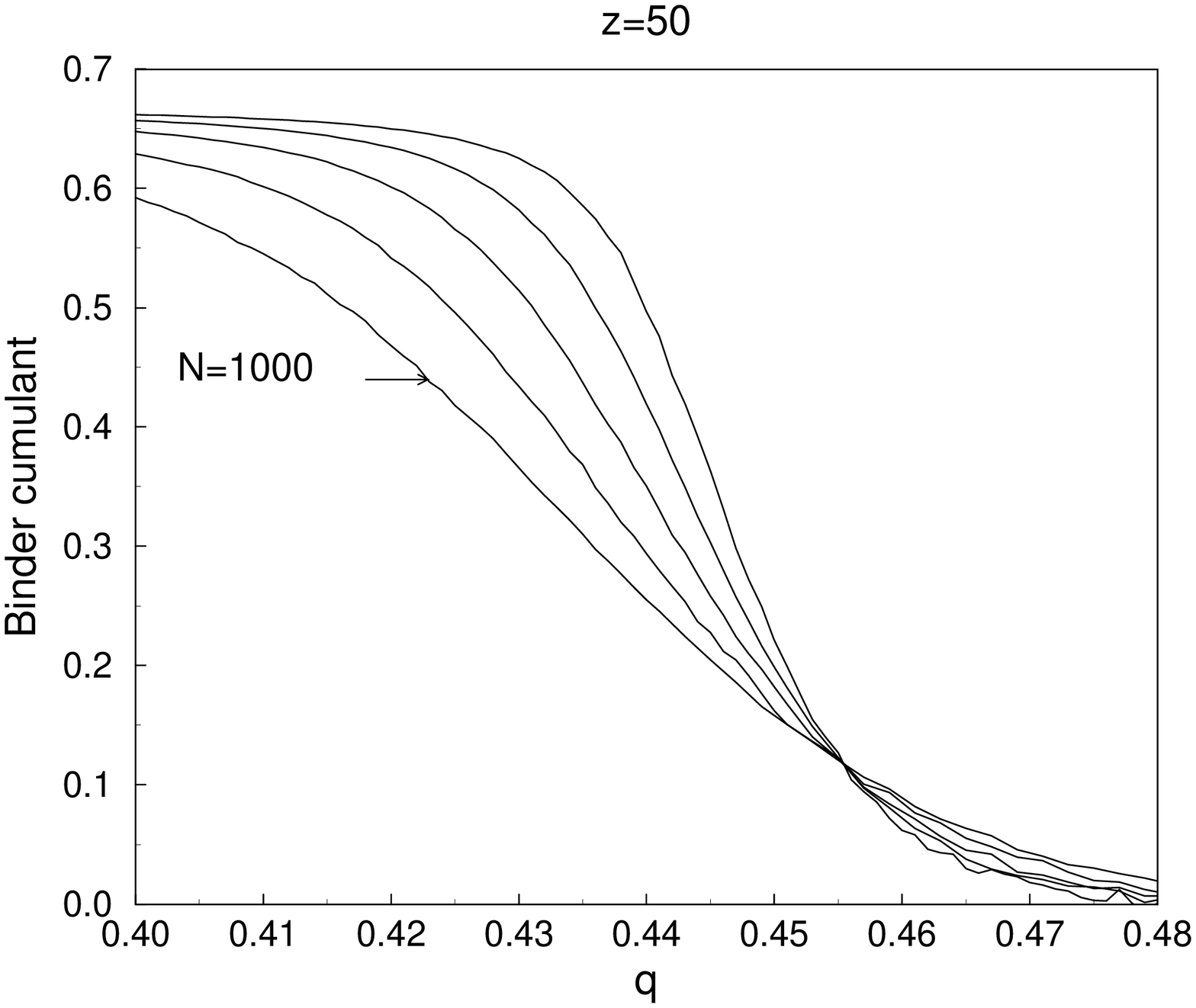}
\end{center}
\caption{
Binder's fourth-order cumulant as a function of $q$. We have $z=2$ and
 $z=50$ for $N=1000$, $2000$, $4000$, $8000$ and $16000$ sites.}
\end{figure}
 
\begin{figure}[hbt]
\begin{center}
\includegraphics[angle=-90,scale=0.60]{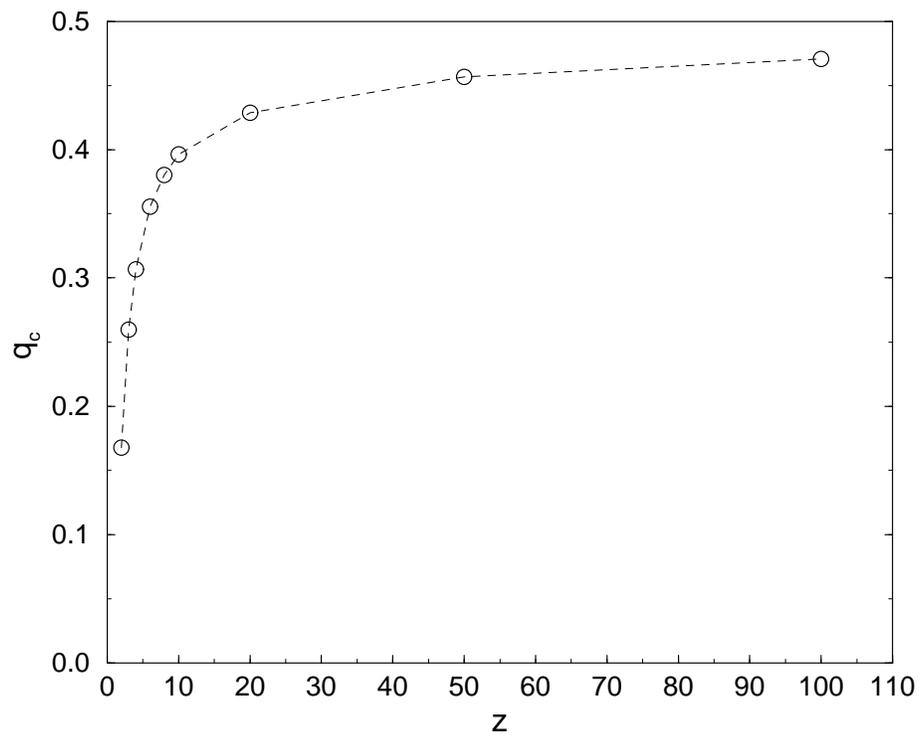}
\end{center}
\caption{The phase diagram, showing the dependence of critical 
noise parameter $q_{c}$ on connectivity $z$.
} 
\end{figure}
 
\begin{figure}[hbt]
\begin{center}
\includegraphics[angle=-90,scale=0.60]{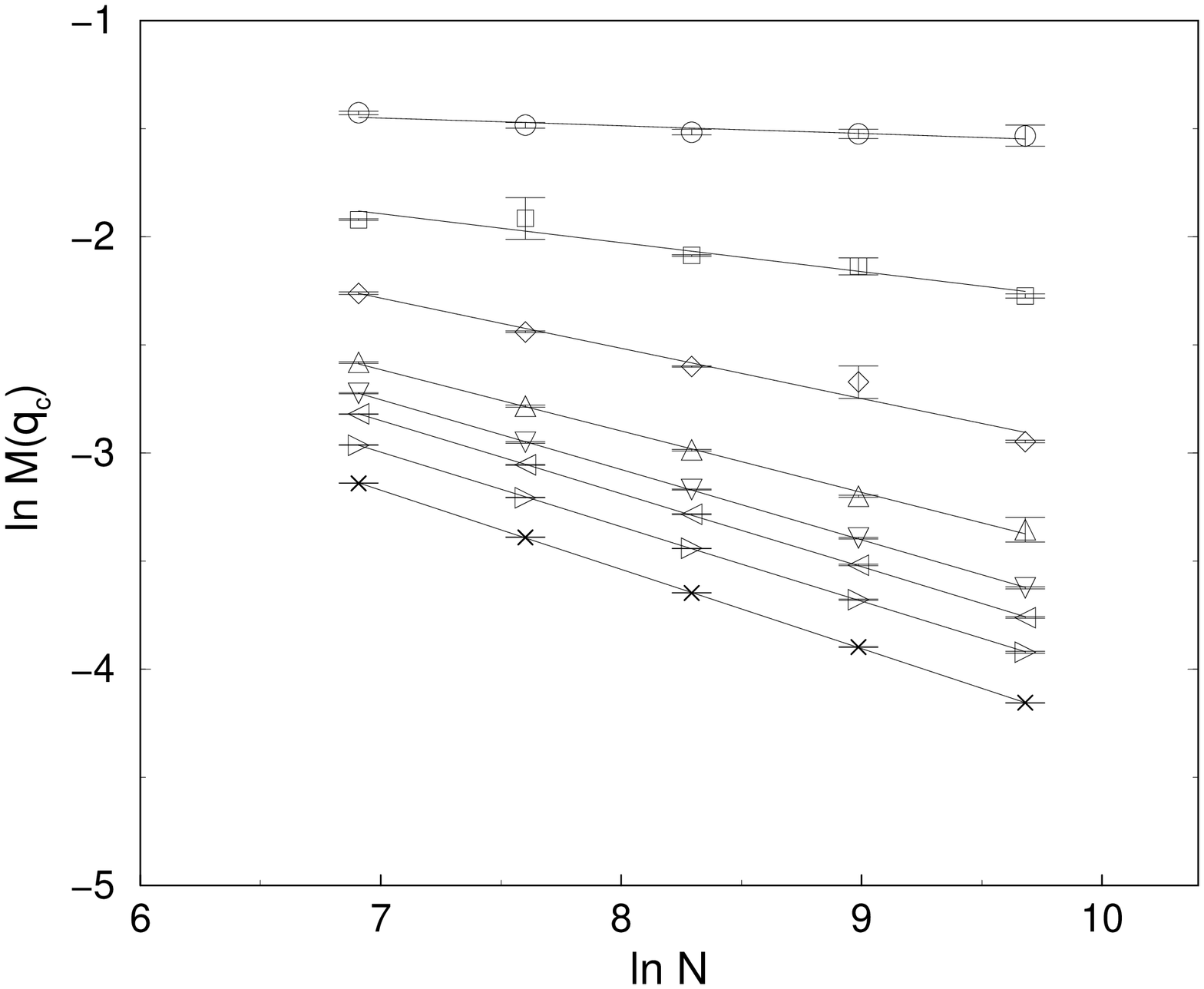}
\end{center}
\caption{ln $M(q_{c})$ versus  ln $N$. From top to bottom, $z=2$, $3$, $4$, $6$,
$8$, $10$, $20$, and $50$. }
\end{figure}

\bigskip

\begin{figure}[hbt]
\begin{center}
\includegraphics[angle=-90,scale=0.60]{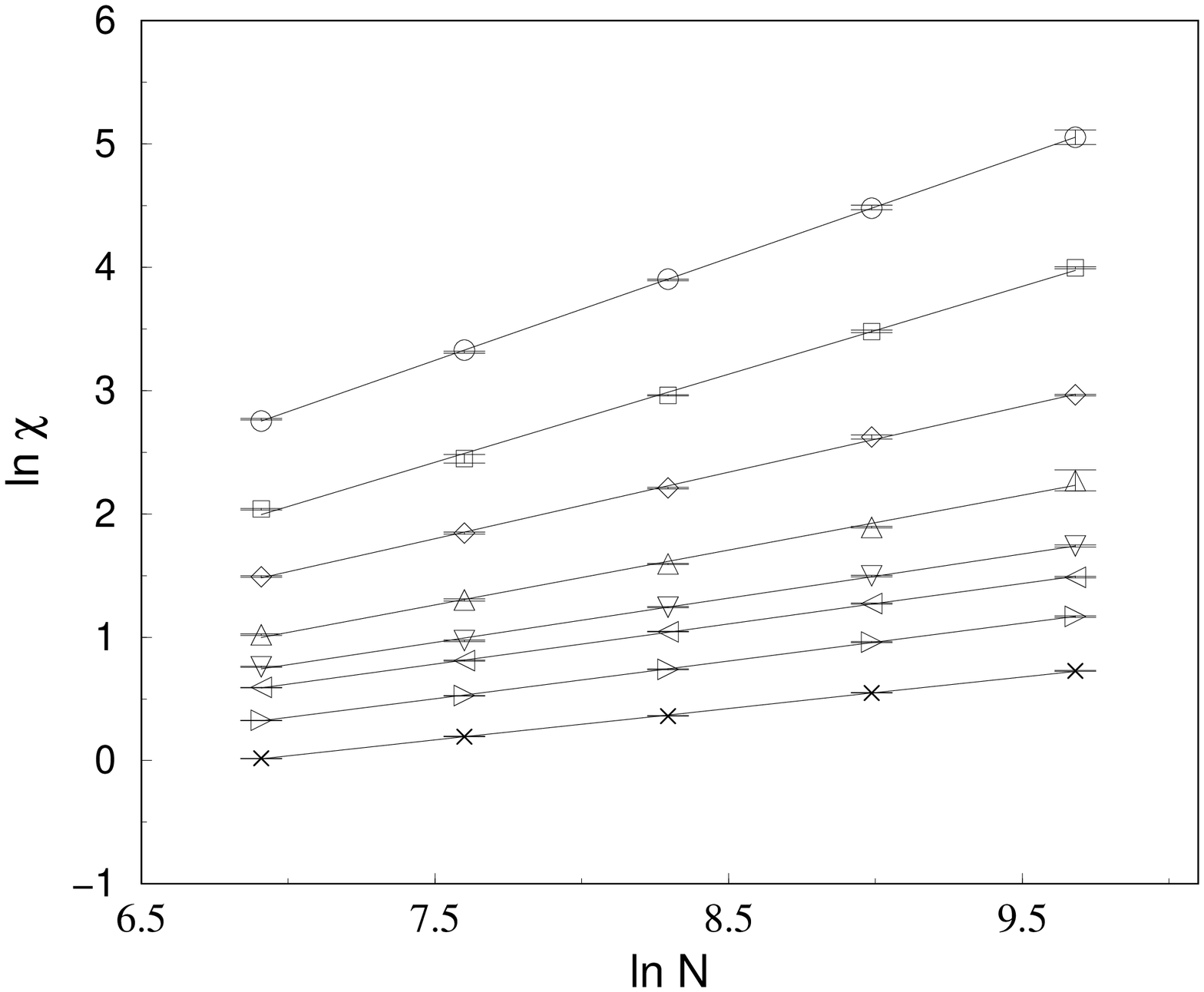}
\end{center}
\caption{ln $\chi$ versus  ln $N$. From top to bottom $z=2$, $3$, $4$, $6$,
 $8$, $10$, $20$, and $50$.
}
\end{figure}
\bigskip
 
\begin{figure}[hbt]
\begin{center}
\includegraphics[angle=-90,scale=0.60]{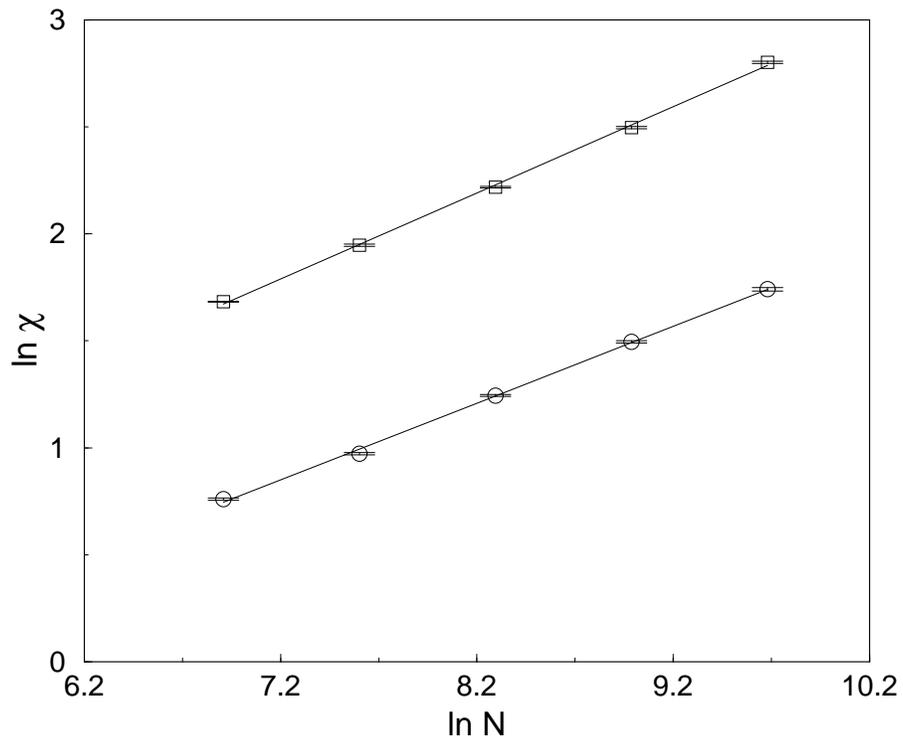}
\end{center}
\caption{ Plot of ln $\chi^{max}(N)$ (square) and ln$ \chi(q_{c})$ (circle) versus ln $N$ for 
connectivity $z=8$.}
\end{figure}

\begin{figure}[hbt]
\begin{center}
\includegraphics[angle=-90,scale=0.60]{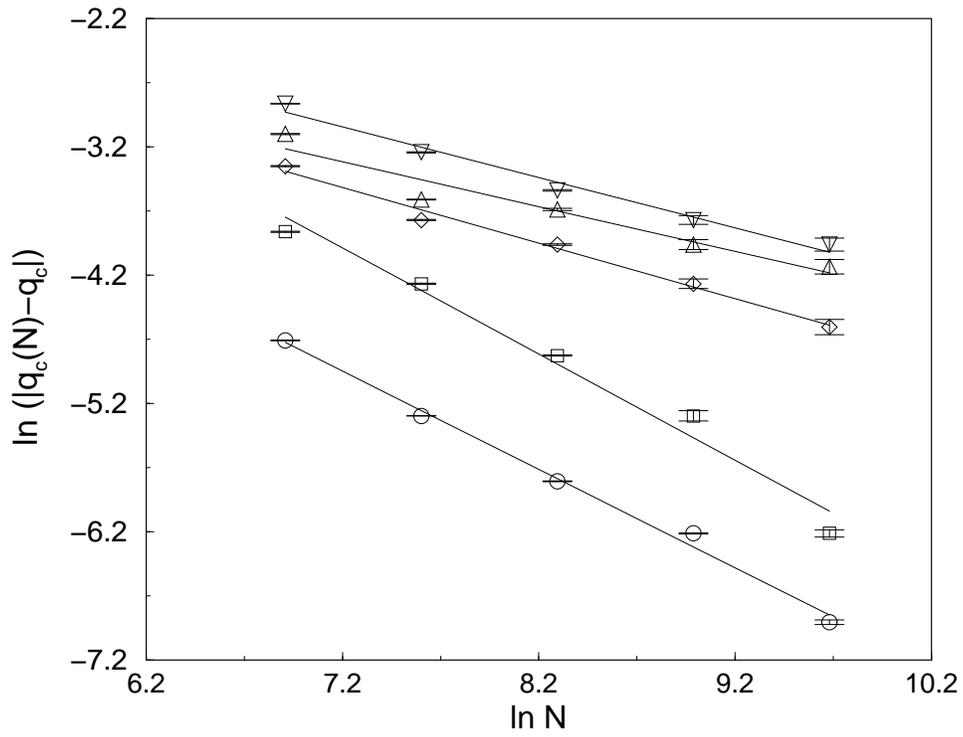}
\end{center}
\caption{ Plot of ln $|q_{c}(N)-q_{c}|$ versus ln $N$. From  bottom to top $z=2$, $3$, $4$, $6$,
 and $8$.}
\end{figure}

\begin{figure}[hbt]
\begin{center}
\includegraphics[angle=-90,scale=0.60]{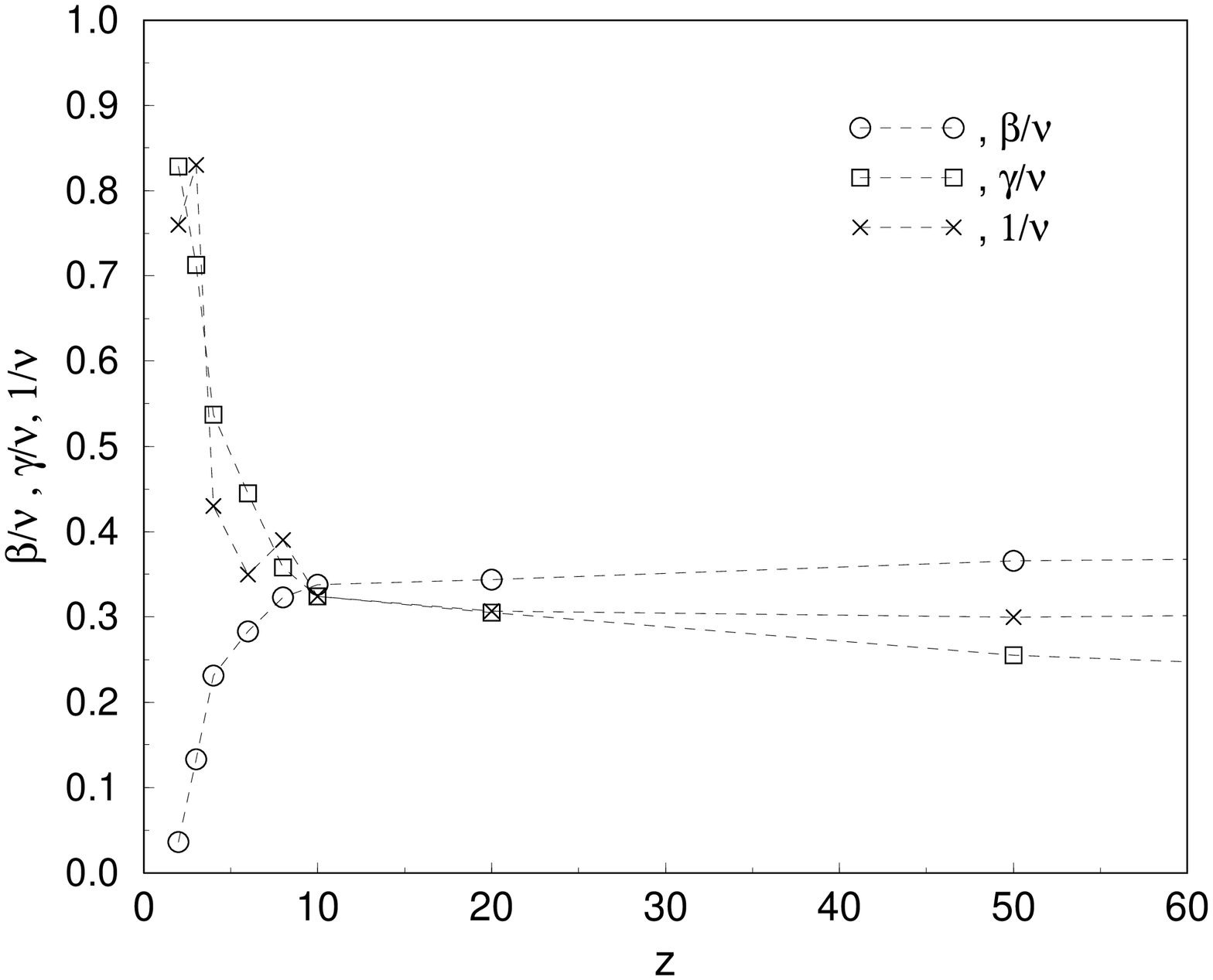}
\end{center}
\caption{ Critical behavior the $\beta/\nu$, $\gamma/\nu$, and $1/\nu$ exponents in function of
connectivity $z$.}
\end{figure}
\bigskip

{\bf Model and Simulaton}

We consider the majority-vote model, on directed 
Barab\'asi-Albert Networks, defined \cite{mario,jff,lima0,brad} by a set of
"voters" or spins variables ${\sigma}$ taking the values $+1$ or
$-1$, situated on every site of an undirected 
Barab\'asi-Albert Networks with $N$ sites, and evolving in time by single spin-flip
like dynamics with a probability $w_{i}$ given by
\begin{equation}
w_{i}(\sigma)=\frac{1}{2}\biggl[ 1-(1-2q)\sigma_{i}S\biggl(\sum_{\delta
=1}^{k_{i}}\sigma_{i+\delta}\biggl)\biggl],
\end{equation}
where $S(x)$ is the sign $\pm 1$ of $x$ if $x\neq0$, $S(x)=0$ if $x=0$, and the 
sum runs over all nearest neighbours of $\sigma_{i}$. In this network, each new site
added to the network selecting $z$,
already existing sites as neighbours influencing it; the newly
added spin does not influence these neighbours. The control parameter
$q$ plays the role of the temperature in equilibrium systems and measures
the probability of aligning antiparallel to the majority of neighbours.

To study the critical behavior of the model we define the variable
$m=\sum_{i=1}^{N}\sigma_{i}/N$. In particular , we were interested in the
magnetisation, susceptibility and the reduced fourth-order cumulant:
\begin{equation}
M(q)=[<|m|>]_{av},
\end{equation}
\begin{equation}
\chi(q)=N[<m^2>-<|m|>^{2}]_{av},
\end{equation}
\begin{equation}
U(q)=\biggl[1-\frac{<m^{4}>}{3<|m|>^{2}}\biggl]_{av},
\end{equation}
where $<...>$ stands for a thermodynamics average and $[...]_{av}$ square brackets
for a averages over the 20 realizations. 

These quantities are functions of the noise parameter $q$ and obey the finite-size
scaling relations

\begin{equation}
M=N^{-\beta/\nu}f_{m}(x)[1+ ...],
\end{equation}
\begin{equation}
\chi=N^{\gamma/\nu}f_{\chi}(x)[1+...],
\end{equation}
\begin{equation}
\frac{dU}{dq}=N^{1/\nu}f_{U}(x)[1+...],
\end{equation}
 where $\nu$, $\beta$, and $\gamma$ are the usual critical 
exponents, $f_{i}(x)$ are the finite size scaling functions with
\begin{equation}
x=(q-q_{c})N^{1/\nu}
\end{equation}
being the scaling variable, and the brackets $[1+...]$ indicate
corretions-to-scaling terms. Therefore, from the size dependence of $M$ and $\chi$
we obtained the exponents $\beta/\nu$ and $\gamma/\nu$, respectively.
The maximum value of susceptibility also scales as $N^{\gamma/\nu}$. Moreover, the
value of $q$ for which $\chi$ has a maximum, $ q_{c}^{\chi_{max}}=q_{c}(N)$,
is expected to scale with the system size as
\begin{equation}
q_{c}(N)=q_{c}+bN^{-1/\nu},
\end{equation}
were the constant $b$ is close to unity. Therefore, the  relations $(7)$ and $(9)$
are used to determine the exponente $1/\nu$. We have checked also if the calculated
exponents satisfy the hyperscaling hypothesis
\begin{equation}
2\beta/\nu+\gamma/\nu=D_{eff}
\end{equation}
in order to get the effective dimensionality, $D_{eff}$, for various values of $z$.

We have performed Monte Carlo simulation on {\it undirected} Barab\'asi-Albert networks with
various values of connectivity $z$. For a given $z$, we used systems
of size $N=1000$, $2000$, $4000$, $8000$, and $16000$. We waited $10000$ Monte Carlo
steps (MCS) to make the system reach the steady state, and the time averages were
estimated from the next $ 10000$ MCS. In our simulations, one MCS is accomplished
after all the $N$ spins are updated. For all sets of parameters, we have generated
$20$ distinct networks, and have simulated $20$
independent runs for each distinct network.

\bigskip

{\bf Results and Discussion}

In Fig. 1 we show the dependence of the magnetisation $M$  and the susceptiblity
$\chi$  on the noise parameter, obtained from simulations on {\it undirected}
Barab\'asi-Albert network with $16000$ sites and several values of connectivity $z$.
In the part (a) each  curve for $M$, for a given value of  $N$ and $z$, suggests
that there is  a phase transition from an ordered state to a disordered state. The
phase transition occurs at a value of the critical noise parameter $q_{c}$, which is
an increasing function the connectivity $z$  of the directed Barab\'asi-Albert
network. In the part (b) we
show the corresponding behavior of the susceptibility $\chi$, the value  of $q$
where  $\chi$ has a maximum is here identified as $q_{c}$. In Fig. 2  we plot
Binder's fourth-order cumulant for different values of $N$ and two different values
of $z$. The critical noise parameter $q_{c}$, for a given value of $z$, is estimated
as the point where the curves for different system sizes $N$ intercept each other.
In Fig 3 the phase diagram is shown as a function of the critical noise
parameter $q_{c}$ on connectivity $z$ obtained from the data of Fig. 2. 

The phase diagram of the majority-vote model on {\it undirected} Barab\'asi-Albert network
shows that for a given network (fixed $z$ ) the system becomes ordered for
$q<q_{c}$, whereas it has zero magnetisation for $q\geq q_{c}$. We notice that the
increase of $q_{c}$ in function the $z$ is slower of the one than in \cite{brad}. In
the Fig. 4 and 5 we plot the dependence of the magnetisation and susceptibility, respectively at $q=q_{c}$ with the system
size. The slopes of curves correspond to the exponent ratio $\beta/\nu$ and $\gamma/\nu$ of according
to Eq. (5) and (6), respectively.
The results show that the exponent ratio $\beta/\nu$ increase and $\gamma/\nu$ decrease at $q_{c}$ when $z$ increase, see Table I.

In Fig. 6 we display the scalings for susceptibility at $q=q_{c}(N)$ (square), $\chi(q_{c}(N))$, and for its maximum amplitude, $\chi_{N}^{max}$, and the scalings for susceptibility at the $q=q_{c}$ obtained from Binder's cumulant, $ \chi(q_{c})$ (circle), versus $N$ for connectivity $z=8$. The exponents ratio $\gamma/\nu$ are obtained from the slopes
of the straight lines. For almost all the values of $z$, the exponents $\gamma/\nu$ of the two estimates agree (along with errors), see Table I. An increased $z$ means a  tendency to decrease the exponent ratio $\gamma/\nu$, see Table I, they agree with the results of Luiz et al \cite{brad}, but disagree with the results of Lima \cite{lima1} for {\it directed} Barab\'asi-Albert network, where the values of the exponents ratio $\gamma/\nu$ are  all different and with a slight tendency to increase at $q_{c}$ and decrease at $q_{c}(N)$. Therefore we can use the Eq. (9) , for fixed $z$, obtain the critical exponent $1/\nu$, see Fig. 7. In the Fig. 8 we show the critical behavior of 
$\beta/\nu$, $\gamma/\nu $ and $1/\nu$ the exponentes in function of connectivity $z$. 

To obtain the critical exponent $1/\nu$, we calculated numerically  
$U^{'}(q)=dU(q)/dq$ at the critical point for each values of $N$ at connectivity fixed $z$. The results
are  well in agreement with the scaling relation (7) . Then , we also can calculate the exponents $1/\nu$, through this relation. Therefore we do not have to get the values of the exponents $1/\nu$ for each connectivity $z$

The Table I summarizes the values of $q_{c}$, the exponents $\beta/\nu$, $\gamma/\nu$,  and 
the effective dimensionality of systems. For all values of $z$, $D_{eff}=1$, which has been obtained from the Eq. (9), therefore when $z$ increases,  $\beta/\nu$ increases and $\gamma/\nu$ decreases at $q_{c}$, thus providing the value of $D_{eff}=1$ (along with errors). Therefore, the {\it undirected} Barab\'asi-Albert network has the same effective dimensionality that Erd\"os-R\'enyi's random graphs \cite{brad} and {\it directed} Barab\'asi-Albert network \cite{lima1}. J. M. Oliveira \cite{mario} showed which majority-vote model  defined on regular lattice has critical exponents that
fall into the same class of universality as the corresponding equilibrium Ising model. Campos et al \cite{campos} investigated the  critical behavior of the majority-vote on small-world networks by rewiring the two-dimensional square lattice, Luiz et al \cite{brad} studied this model on Erd\"os-R\'enyi's random graphs, and Lima et al \cite{lima0} also studied this model on Voronoy-Delaunay
lattice and Lima {\it directed} Barab\'asi-Albert network \cite{lima1}. The results obtained these authors show that the critical exponents of majority-vote model belong to different universality classes.

\begin{table}[h]
\begin{center}
\begin{tabular}{|c c c c c c c|}
\hline
\hline
$ z $ & $q_{c}$ & $\beta/\nu$& ${\gamma/\nu}^{q_{c}}$ & ${\gamma/\nu}^{q_{c}(N)}$ & 
 $ 1/\nu$ & $ D_{eff}$\\
\hline
$ 2 $ & $ 0.167(3) $ & $ 0.036(8) $ & $ 0.828(6) $ & $ 0.805(11) $ & $ 0.76(3) $ & $ 0.90(1)$\\

$ 3 $ & $ 0.259(2) $ & $ 0.133(21) $ & $ 0.713(18) $ & $ 0.655(31) $ & $ 0.83(7) $ & $ 0.979(27)$\\ 
$ 4 $ & $ 0.306(3) $ & $ 0.231(22) $ & $ 0.537(8) $ & $ 0.519(17) $ & $ 0.43(2) $ & $ 0.999(23)$\\
$ 6 $ & $ 0.355(2) $ & $ 0.283(8) $ & $ 0.445(15) $ & $ 0.423(3) $ & $ 0.35(5) $ & $ 1.011(17)$\\
$ 8 $ & $ 0.380(6) $ & $ 0.323(2) $ & $ 0.358(7) $ & $ 0.405(6) $ & $ 0.39(3) $ & $ 1.004(7)$\\
$ 10 $ & $ 0.396(3) $ & $ 0.338(2) $ & $ 0.324(2) $ & $ 0.380(3) $ & $ 0.324(5) $ & $ 1.000(2)$\\
$ 20 $ & $ 0.428(2) $ & $ 0.344(2) $ & $ 0.305(2) $ & $ 0.350(2) $ & $ 0.307(5) $ & $ 0.993(2)$\\
$ 50 $ & $ 0.456(3) $ & $ 0.366(2) $ & $ 0.255(2) $ & $ 0.341(3) $ & $ 0.30(1) $ & $ 0.987(2)$\\
$ 100 $ & $ 0.471(3) $ & $ 0.373(2) $ & $ 0.218(5) $ & $ 0.330(3) $ & $ 0.308(4) $ & $ 0.964(5)$\\
\hline
\hline
\end{tabular}
\end{center}
\caption{ The critical noise $q_{c}$, the critical exponents, and the effective dimensionality $D_{eff}$
, for  undirected Barab\'asi-Albert network with connectivity $z$. Error bars are statistical only.} \label{table1}
\end{table}

Finally, we remark that our MC results obtained on {\it undirected} Barab\'asi-Albert network for majority-vote model show that critical exponents are different from the results of 
\cite{mario} for regular lattice, of Luiz et al \cite{brad} for Erd\"os-R\'enyi's random graphs and Lima \cite{lima1}.

\bigskip
 
{\bf Conclusion}
 
In conclusion, we have presented a very simple nonequilibrium model on
{\it undirected} Barab\'asi-Albert network \cite{sumour,sumourss}. Different of 
of Ising model, in these networks, the majority-vote model presents a
second-order phase transition which occurs with 
connectivity $z>1$. The exponents obtained are different from the other models. 
Nevertheless, our Monte Carlo simulations have demonstrated that the 
effective dimensionality $D_{eff}$ equals unity, for all values of $z$,
which agrees with the results de Luiz et al \cite{brad}.
However, when $z$ grows, the exponents in the critical point $q_{c}$, $\beta/\nu$
obtained through Binder's cumulant grows and exponents $\gamma/\nu$  decrease, satisfying
the hyperscaling relation with $D_{eff}= 1$.

  The author thanks Wagner Figueiredo for help with linux cluster  from
UFSC (Santa Catarina-Brazil), and D. Stauffer for many suggestions and fruitful
discussions during the development this work and also for the revision of
this paper. I also acknowledge the Brazilian agency FAPEPI
(Teresina-Piau\'{\i}-Brasil) for  its financial support. This work also was supported the
system SGI Altix 1350 the computational park CENAPAD.UNICAMP-USP, SP-BRAZIL.

\end{document}